\documentclass[aps, twocolumn,noshowpacs,preprintnumbers,amsmath,amssymb]{revtex4}

\usepackage{verbatim}
\usepackage{wasysym}

\usepackage{graphicx}
\usepackage{dcolumn}
\usepackage{bm}
\usepackage{wrapfig}

\begin{document}

\title{Radial elasticity of multi-walled carbon nanotubes}

\author{Ismael Palaci}
\affiliation{School of Physics, Georgia Institute of Technology,\\ 837 State Street, Atlanta, Georgia 30332, USA}
\affiliation{Institute of the Physics of Nanostructures, Ecole Polytechnique Federale de Lausanne (EPFL),\\ CH-1015 Lausanne, Switzerland}
\author{Stephan Fedrigo}
\affiliation{Institute of the Physics of Nanostructures, Ecole Polytechnique Federale de Lausanne (EPFL),\\ CH-1015 Lausanne, Switzerland}
\author{Harald Brune}
\affiliation{Institute of the Physics of Nanostructures, Ecole Polytechnique Federale de Lausanne (EPFL),\\ CH-1015 Lausanne, Switzerland}
\author{Christian Klinke}
\affiliation{Institute of the Physics of Nanostructures, Ecole Polytechnique Federale de Lausanne (EPFL),\\ CH-1015 Lausanne, Switzerland}
\affiliation{IBM Watson Research Center, 1101 Kitchawan Road,\\ Yorktown Heights, New York 10598, USA}
\author{Michael Chen}
\affiliation{School of Physics, Georgia Institute of Technology,\\ 837 State Street, Atlanta, Georgia 30332, USA}
\author{Elisa Riedo}
\affiliation{School of Physics, Georgia Institute of Technology,\\ 837 State Street, Atlanta, Georgia 30332, USA}
\email{elisa.riedo@physics.gatech.edu}

\begin{abstract} 

We report an experimental and a theoretical study of the radial elasticity of multi-walled carbon nanotubes as a function of external radius. We use atomic force microscopy and apply small indentation amplitudes in order to stay in the linear elasticity regime. The number of layers for a given tube radius is inferred from transmission electron microscopy, revealing constant ratios of external to internal radii. This enables a comparison with molecular dynamics results, which also shed some light onto the applicability of Hertz theory in this context. Using this theory, we find a radial Young modulus strongly decreasing with increasing radius and reaching an asymptotic value of 30 $\pm$ 10 GPa.

\end{abstract}

\maketitle

The exceptional mechanical, electrical, and thermal properties \cite{1,2,3,4,5,6,7,8} of carbon nanotubes (CNTs) have attracted great scientific and technological interest. CNTs have cylindrical symmetry with axial mechanical properties characterized by the strong in plane covalent C-C bond. The strength of this bond gives rise to an extraordinary axial stiffness, as pointed out by several experimental \cite{1,9,10} and theoretical studies \cite{7,8,11} finding values for the axial Young modulus of about 1 TPa. In graphite, the $C_{11}$ in plane elastic constant is 1.06 TPa, while the perpendicular elastic constant $C_{33}$ is only 36 GPa \cite{7}. Similarly the radial Young modulus of CNTs is expected to be much smaller than the axial one. Evidence for the softness of CNTs in the radial direction has been reported in experiments under hydrostatic pressure \cite{12}, where a critical pressure of only 2 GPa has led to the collapse of single-walled CNTs with a radius of 0.7 nm. Achieving a fundamental understanding of the radial deformability of CNTs is important for applying them in nanoelectromechanical and nanoelectronic systems. For example, the radial deformation of CNTs may strongly affect their electrical properties \cite{3,13,14,15,16}. However, our quantitative
understanding of the radial elasticity of CNTs is so far based on studies performed on only one tube, with an unknown number of layers, and using deformations up to
the nonlinear regime \cite{17,18,19,20}. 

In principle, the simplest way to measure the radial elasticity of CNTs would be to indent an atomic force microscope (AFM) tip into a NT adsorbed at a surface and to measure force vs. indentation curves. However, in practice, such measurements are very challenging, since in order to stay in the linear elastic regime, one has to measure forces of a few nanonewtons vs displacements of a few \AA{}. Some authors have proposed an alternative AFM based method to investigate the radial elasticity of CNTs \cite{17}. While scanning the tip across the sample, the authors vertically vibrate the cantilever in noncontact or tapping mode with amplitudes in the range of several hundreds of \AA{} and with the turning point situated a few \AA{} above the sample. Because of the large amplitudes, a considerable fraction of the signal arises from the van der Waals forces acting between the tip and the tube, and only a small part comes from the elastic properties of the tube. Therefore in these experiments in order to extract quantitative results on the radial deformation of a CNT it is necessary to evaluate the van der Waals forces taking the cantilever, tip, and sample geometry into account, which is far from trivial \cite{21}.

Here, we present quantitative measurements of the radial elasticity of 39 multi-walled CNTs with external radii ranging from 0.2 to 12 nm and having a constant ratio of external to internal radii of $R_{ext}/R_{int} = 2.2 \pm 0.2$. We underline that the NT with $R_{ext} = 0.2$ is most likely a single-walled NT. By means of modulated nanoindentation with an AFM \cite{22}, we find that the radial stiffness strongly increases with decreasing external diameter. The radial Young modulus $E_{rad}$ is extracted from the experimental results by applying the Hertz model. $E_{rad}$ is found to decrease to an asymptotic value of 30 GPa for larger tube
sizes. We also perform molecular dynamics (MD) simulations with empirical C-C potentials to mimic the experiments. Force-indentation curves obtained by the simulations indicate a similar trend in $E_{rad}(R_{ext})$.

The multi-walled CNTs are produced by chemical vapor deposition (CVD) using acetylene as carbon feedstock \cite{23}. A drop of an alcohol suspension of the obtained
CNTs is deposited onto a silicon surface and the solvent is allowed to evaporate at room temperature. In this way, the CNTs are adsorbed on the Si substrate with their principal axis parallel to it. The CVD production method generates CNTs with $R_{ext}/R_{int} = 2.2 \pm 0.2$, as obtained from a large number of measurements with transmission electron microscopy (TEM) (see inset of Fig. 1). The morphology and the mechanical properties of CNTs have been measured with an AFM \cite{24} operating in contact mode in ambient conditions and equipped with commercial SiN cantilevers with a tip radius of typically 35 nm. Normal cantilever spring constants, klev, have been carefully calibrated and typical values were about 46 N/m. For each NT, the tip radius has been explicitly determined in two ways: first, by using the equation $R_{tip} = w^{2} / 16R_{NT}$, where $R_{NT}$ is the tube radius inferred from its apparent height and $w$ is its apparent width; second, by imaging the tip with a scanning electron microscope. Both methods yielded consistent results.

\begin{figure}[htbp]
  \centering
  \includegraphics[width=0.45\textwidth]{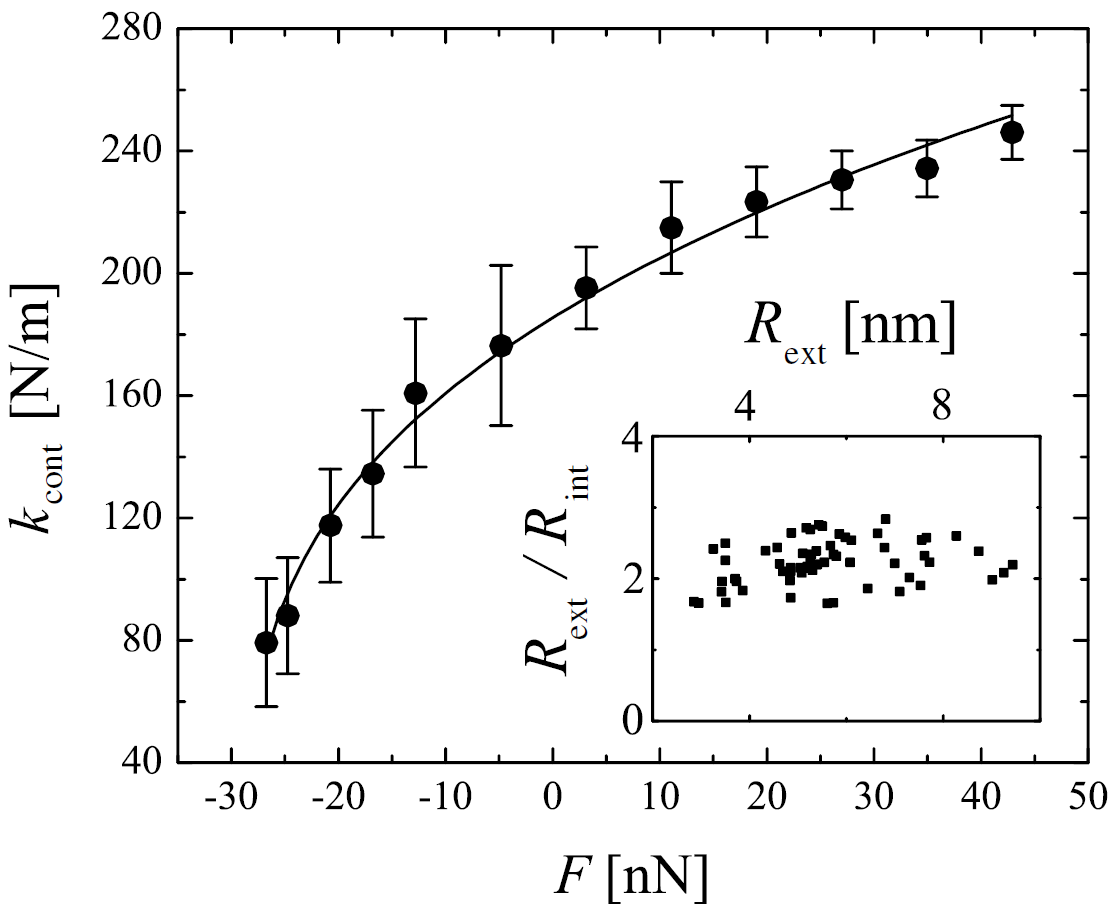}
  \caption{\textit{Experimental normal contact stiffness vs. normal indentation force $F$ for a 3 nm tube radius. Errors presented here are due to mean errors on the detection signal $dF$ \cite{28}, taking into account the uncertainty on the cantilever stiffness. Experimental data are fitted with Eqs. (1) and (2). In the inset
we show $R_{ext}/R_{int}$ as a function of $R_{ext}$, as obtained by TEM.}}
\end{figure}

MD simulations are performed by modeling the AFM tip as a rigid continuous sphere and the NT by atoms interacting through an empirical potential. Forces between carbon atoms are derived from a two-body pair energy plus a three-body angular penalty for the covalent energy (intralayer energy), as developed by Marks \cite{25}], and from a truncated Lennard-Jones potential for the interlayer energy, as applied by Lu \cite{7}. The free potential parameters are fitted on the bulk graphite elastic constants, $C_{11}$, $C_{12}$, and $C_{33}$, the cohesive energy, and the two lattice constants. CNTs are built with graphene sheets spaced by an inter-wall
distance as close as possible to the graphite interlayer distance, the chirality being a free parameter. Subsequently, the CNTs are compressed between the rigid sphere and a rigid plane using short range, purely repulsive potentials for both interactions. The two ends of the NTs are frozen. The NT length and the sphere radius are, respectively, fixed at 20 and 12 nm. In all cases studied, the largest diameter of the contact area is smaller than 1.4 nm. Technically, the sphere is slowly moved against the NT, while the kinetic energy is periodically removed. Expressed in the usual MD units (m.a.u., eV, and \AA{}), the time step is 0.4 and the sphere velocity is equal to or lower than $2 \cdot 10^{-5}$ \AA{} per time step.

Normal modulated nanoindentation consists of indenting an AFM tip in a sample up to a fixed distance while small oscillations are applied to the sample. Oscillations
and indentation are colinear, normal to the substrate and to the NT long axis. The amplitude of the oscillations is chosen very small, 1.3 \AA{} in our case, in order to remain in the sticking regime. In this amplitude range and experimental geometry, the normal force $F$ required to move vertically the NT's substrate by a distance $D$ with respect to the cantilever support coincides with the force needed to elastically stretch two springs in series \cite{26,27}: the cantilever, with stiffness $k_{lev}$, and the tip-sample contact, with stiffness $k_{cont}$. If $D$ is the total normal displacement of the NT's substrate, i.e., $D$ is equal to cantilever bending plus tip and NT normal deformation, and $F$ is the total normal force, this configuration allows the measurement of the total stiffness $k_{tot}$ at each load, defined by the relation

\begin{equation}
  dF/dD = k_{tot} = (1/k_{lev} + 1/k_{cont})^{-1}
\end{equation}

Since $k_{lev}$ is known, a measurement of $dF/dD$ at different normal loads leads to the value of $k_{cont}$ as a function of $F$ \cite{28}. Figure 1 shows the results of the measurement of $k_{cont}(F)$ for a NT with a radius of 3 nm. $F = 0$ nN corresponds to the cantilever being unbent. The fact that the tip and the sample remain in contact at negative external loads indicates the presence of an adhesive force.

By integrating the equation $dF = k_{cont} \cdot dz$, where $z$ is the indentation of the tip in the NT \cite{29}, we obtain $F$ vs. indentation $z$ from the experimental curves $k_{cont}(F)$. The result is shown in Fig. 2(a) for NT radii from 0.2 to 5.25 nm. If we call $F(z)/z$ the radial stiffness of the NT, Fig. 2(a) indicates that the radial stiffness increases when the tube radius is decreasing for any value of $z$ in the range explored by the experiment. Figure 2(b) presents the normal force vs. the indentation distance obtained by the simulations and with $R_{ext}/R_{int}$ kept close to the experimental value for CNTs with 2 to 6 layers. The respective NT external radii are 0.61, 1.22, 1.82, 2.43, and 3.65 nm, while the ratios between external and internal radii are kept constant and equal to 2.2. In agreement with experiment, the nanotubes radial stiffness also increases when its radius decreases, and again in agreement this effect is less pronounced
for larger NT radii. Compared to the results of Fig. 2(a), normal forces at equivalent indentation distances are typically 1 order of magnitude lower in the simulation. This is mainly attributed to the fact that the tip radius in the experiment is a factor of 2 larger than the sphere radius in the simulation.

We can extract the radial CNT Young modulus from $k_{cont}$ vs. $F$ measurements by modeling the contact between the AFM tip and the CNTwith the Hertz model \cite{13,17,18}. We underline that the extracted $E_{rad}$ is therefore the radial linear elasticity. The Hertz model predicts a 3/2 power law dependence of $F$ on $z$, which we indeed observe in our experiments [see Fig. 2(a)]. From the calculations, we find the 3/2 power law for filled CNTs, while there are deviations from $F \propto z^{3/2}$ for hollow NTs with the cross sections used in experiment [$R_{ext}/R_{int} = 2.2$; see Fig. 2(b)]. We attribute the deviations to the fact that the calculations have been performed for technical reasons with a smaller tip radius than the one in the experiment. This suggests that our experiment is just at the limit where the Hertz model might be applied, whereas the size of the contact in the simulations falls below this limit.

\begin{figure}[htbp]
  \centering
  \includegraphics[width=0.45\textwidth]{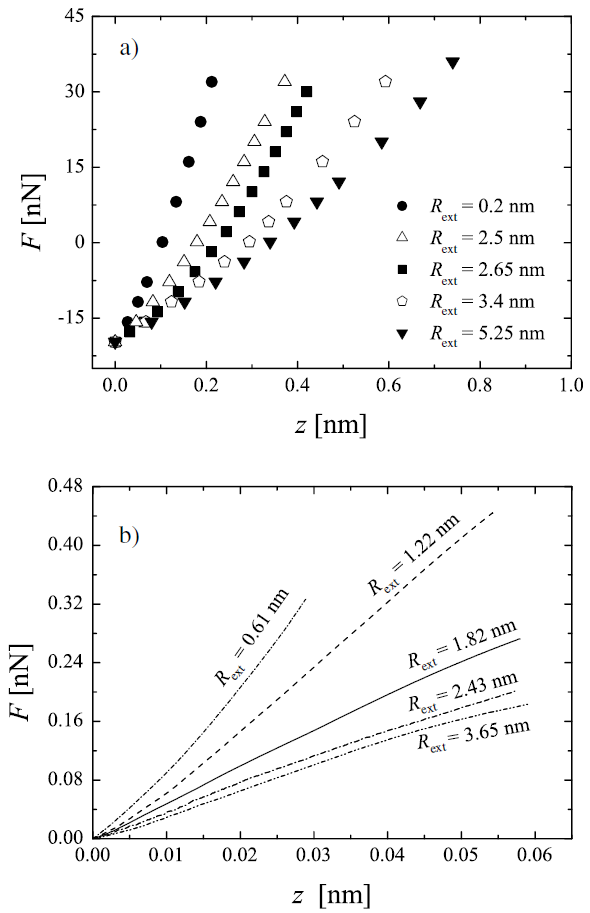}
  \caption{\textit{(a) Normal force as a function of indentation for NTs of different radii, obtained by the integration of experimental $1/k_{cont}$ vs. $F$ curves using the trapeze method. (b) Theoretical normal force as a function of indentation for NTs of different radii (different number of layers), obtained by simulating the indentation of a rigid sphere in a NT.}}
\end{figure}

Under the assumptions of standard elasticity theory, the Hertz model gives the dependence of the indentation distance $z$ vs. the normal force $F$ between two elastic solids in contact \cite{30}. Although very sophisticated extensions of this model were developed to include the effect of the adhesion at low external forces \cite{31}, in the context of this work it suffices to use the first level approximation, consisting of an additive correction of the normal force $F$. We consider the contact between a sphere and a cylinder (corresponding to the tip and the NT), and we include the adhesive force $F_{adh}$, which is experimentally determined. The Hertz theory gives

\begin{equation}
  k_{cont} = \beta \left( \frac{R(F + F_{adh}}{\tilde{K}^{2}} \right)^{1/3}
\end{equation}

with $1/R = 1/R_{tip} + 1/2R_{NT}$ and $\tilde{K} = 3/4 \cdot ( (1-\nu_{1}^{2})/E_{1} + (1-\nu_{2}^{2})/E_{2} )$, where $\nu_{1,2}$ and $E_{1,2}$  are, respectively, the Poisson ratios and radial Young moduli of the tip and NT. $\beta$ is a coefficient that takes the geometrical aspect of the contact area into account \cite{32}. $k_{cont}$ vs. $F$ (as in Fig. 1) is then fitted with Eq. (2), the Young modulus $E_{2} = E_{rad}$  being the only free fit parameter for each NT. The elastic constants of the SiN tip are $\nu_{1} = 0.27$ and $E_{1} = 155$ GPa \cite{33}. The Poisson ratio of the NT is taken as $\nu_{2} = 0.28$, a mean value of common materials. Obviously, any reasonable errors on $\nu_{2}$ would have only a minor impact on the extracted $E_{rad}$ and even less so on the variation of the modulus with the NT radius.

\begin{figure}[htbp]
  \centering
  \includegraphics[width=0.45\textwidth]{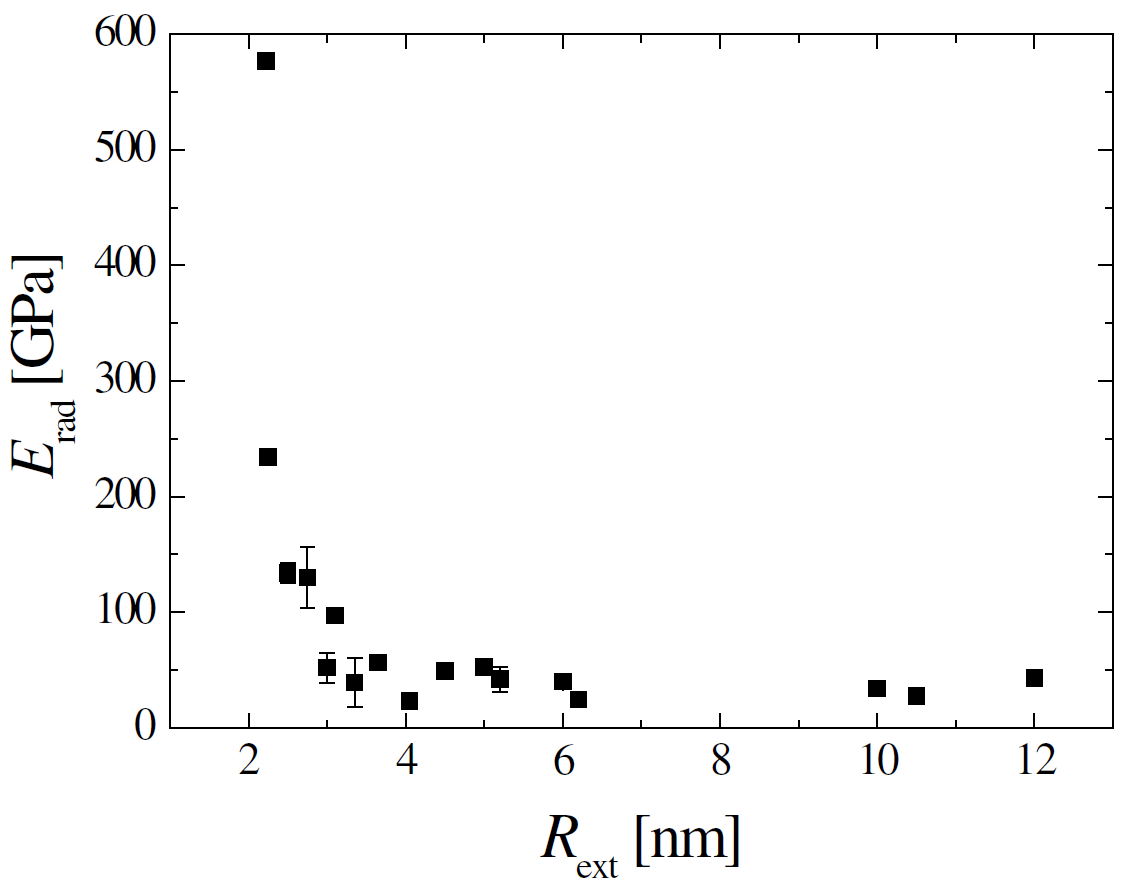}
  \caption{\textit{Experimental values of the radial Young modulus of CNTs as a function of $R_{ext}$ as obtained from normal modulation experiments. Error bars correspond to mean errors on tubes of the same diameter. Errors due to the fit of $k_{cont}$ vs. $F$ are included in the symbol size.}}
\end{figure}

The obtained values of $E_{rad}$ as a function of the CNT's external radius are reported in Fig. 3. The radial Young modulus, as previously observed for the radial stiffness, increases when decreasing the NT radius. More precisely, $E_{rad}$ increases sharply for $R_{ext}$ smaller than 4 nm, while it is almost constant and equal to about 30 $\pm$ 10 GPa for $R_{ext}$ between 4 and 12 nm. This last value is, within the experimental error, equal to the Young modulus of graphite along
its $c$ axis, $E_{graphite} = 36$ GPa \cite{34}. For the NTs studied in this work, $R_{ext}$ is proportional to $R_{int}$ and both are proportional to the number of layers since the distance between layers is approximately constant \cite{7}. Thinking of the elastic energy necessary to enroll a plane, we could deduce that the radial rigidity and hence $E_{rad}$ of a NT should increase by increasing the number of layers and by decreasing the internal radius. This is confirmed by measurements of radial deformations of NTs due to van der Waals forces between the tube and the substrate \cite{35,36}. In both these studies, the radial deformation increases with the radius for single-walled NTs and decreases with the number of layers. Our experiments show that, for small $R_{int}$, $E_{rad}$ increases sharply by decreasing $R_{int}$; we conclude that in this size range the radial rigidity is controlled by the magnitude of $R_{int}$, whereas the number of layers plays a
minor role. This result is in agreement with a previous theoretical study \cite{7} that shows that the elastic properties of a NT with $R_{int} = 0.34$ nm do not change by increasing the number of layers as long as the interlayer distance is fixed to 0.34 nm, i.e., the distance between planes in graphite. A similar finding is also obtained in the simulations of Ref. \cite{36}, where the radial deformation of a single-walled NT is the same of a multi-walled NT when the radius of the first one is equal to $R_{int}$ of the second one. For large $R_{int}$, our experiments show that $E_{rad}$ is almost constant. This could mean that the effect due to the increase of $R_{int}$ is counterbalanced by the increase of the number of layers, up to the point at which the NT's properties reach asymptotically those of graphite. We believe that the behavior shown in Fig. 3 is not restricted to NTs with $R_{ext}/R_{int} = 2$, but it is expected for other ratios larger than 1 since the asymptotic value corresponds to $E$ of graphite.

The radial stiffness of multi-walled CNTs has been investigated experimentally by Yu et al. \cite{17} and by Shen et al. \cite{18}. In both cases, one NT with an unknown number of layers is compressed, the maximum indentation distance being larger than 40\% of the initial diameter. In Ref. \cite{17}, the force vs. indentation distance curves are obtained through a model of the tip-NT van der Waals forces. By interpreting these curves with the Hertz model, they find, for a NT with a diameter of 8 nm, a radial Young modulus between 0.3 and 4 GPa, which is roughly 1 order of magnitude lower than our results for NTs of similar diameters. This discrepancy can be ascribed to a difference in the number of graphene layers forming the NT, which is plausible since the NT preparation techniques are different.
The radial elastic modulus of the NTs obtained in Ref. \cite{18}, where the tubes are deformed up to the nonlinear regime, is hardly comparable to our findings
since its definition differs notably from the one exposed above.

In summary, we measured the radial stiffness and Young modulus of carbon nanotubes. They steeply decline with increasing radii, until the Young modulus takes on an
asymptotic value of 30 $\pm$ 10 GPa for CNTs with $R_{ext} > 5$ nm. The experiments were performed with modulated nanoindentation and on statistically significant amounts of CNTs with well-defined external to internal radii. This trend is very well reproduced by MD simulations.


\begin{thebibliography}{00}
\bibitem{28} A lock-in amplifier is used to vertically modulate the sample position $dD$ and measure $dF$. We have chosen $dD = 1.3$ \AA{} and a frequency of 1.8967 kHz.
\bibitem{29} Tip deformations can be neglected due to the large difference in stiffness, one exception being presumably the smallest tube.
\bibitem{32} $\beta = (\pi / 2k)^{2/3} [ E ( \sqrt{1-k^{2}} )^{1/3} / K ( \sqrt{1-k^{2}} ) ]$, with $k$ a parameter taking into account the elliptic shape of the contact \cite{30} determined by solving the expression $1 + R_{tip}/R_{NT} = [ (1/k^{2} E ( \sqrt{1 - k^{2}} ) - K ( \sqrt{1 - k^{2}} ) ] / [ K ( \sqrt{1 - k^{2}}) - E ( \sqrt{1 - k^{2}})  ]$. $E(k)$ and $K(k)$ are the complete integral of the second kind of type elliptic E and elliptic K defined by $E(k) = \int_{0}^{\pi/2} (1- k^{2} \sin{\theta})^{1/2} d\theta$ and $K(k) = \int_{0}^{\pi/2} (1- k^{2} \sin{\theta})^{-1/2} d\theta$.
\end{thebibliography}
\end{document}